\documentstyle[12pt]{article}
\input{epsfig.sty}
\newcommand{\beq}{\begin{eqnarray}}
\newcommand{\eeq}{\end{eqnarray}}
\begin{document}
\title{Review of QCD, QGP, Heavy Quark Meson Production Enhancement and 
Suppression}
\author{Leonard S. Kisslinger$^{1}$\\
Department of Physics, Carnegie Mellon University, Pittsburgh, PA 15213}
\date{}
\maketitle
\vspace{-1cm}
\noindent
1) kissling$@$andrew.cmu.edu 

\begin{abstract}
  This review of the Quantum Chromodynamics (QCD),the early universe 
Cosmoloical Phase Transition from the  Quark-Gluon Plasma (QGP) to our 
present universe (QCDPT),  Relativistic Heavy Ion Collisions (RHIC) which can 
produce the QGP, the possible detection of the QGP produced by the production 
of mixed hybrid heavy quark mesons. We also review recent studies of the 
production of mixed heavy quark hybrids via RHIC and heavy quark meson 
supression in p-Pb and Pb-Pb collisions.
\end{abstract}


\noindent
mixed hybrid heavy quark mesons

\vspace{2mm}
\noindent
PACS Indices:12.38.Aw,13.60.Le,14.40.Lb,14.40Nd

\section{Introduction}

  This Review of Quantum Chromodynamics (QCD), Quark-Gluon Plasma (QGP), and 
RelativisticHeavy Ion Collisions (RHIC) is based in part on our recent 
review\cite{kd16}. 

We first briefly discuss quarks and gluons, the elementary
particles of QCD and components of the QGP. Next Cosmological Phase
Transitions and the QCD Phase Transition (QCDPT), with the quark condensate
the latent heat, are reviewed. Then the magnetic wall created by bubble 
collisions during the first-order QCDPT, and how this could lead to 
correlations in the Cosmic Microwave Background Radiation, is discussed.
The contrast between predictions of CMBR Correlations using standard QCD and 
predictions using string theory or inflationary models is presented.

The theory of mixed normal and hybrid heavy quark states for charmonium and 
bottomonium mesons and the recent research showing that these mixed hybrid 
states are produced with enhanced cross sections compared to normal meson 
states in RHIC is reviewed.
 
 The possibility that experimental measurements of the production of
mixed heavy quark hybrid states via RHIC could prove the production of the
early universe QGP  via heavy atomic nuclear collisions  with sufficient
energy to reach the temperature of the universe at a time about one millionth
of a second after the Big Bang is briefly discussed.

  Finally, recent publications on the suppression of these mixed hybrid 
state in p-Pb and Pb-Pb collisions are reviewed.
\newpage

\section{A Review of Quantum Chromodynamics (QCD) }

Quantum Chromodynamics (QCD) is the theory of strong interactions. The basic
elementary particles are quarks, fermions with quantum spin 1/2, and gluons,
boson with quantum spin 1. 

The quark quantum field is $q_f$, with $f$ the flavor. The quark flavors are 
$q_f:u,d,s,c,b,t$=up, down, strange, charm, bottom, and top quarks. The most 
important quarks flavors for the present review are the charm and bottom quarks.
The quarks have three colors. The heavy quark masses, which are important for
this review, are  $m_c\simeq$ 1.5 GeV and $m_b \simeq$ 5.0 GeV.

$A_\mu^a$ is the strong interaction field, called the gluon field, with the 
quanta called gluons, and $a$ is the color with gluons having 8 colors.
The interaction of quarks with gluons is 
\beq
\label{qgnteraction}
    V_{qg}&=& \sum_{f} \bar{q}_f \gamma^\mu g_s A_\mu q_{f} \\
         A_\mu&=& \sum_{1}^{8} A_\mu^a \lambda^a/2 \nonumber \; , 
\eeq 
where the quanta of $A_\mu$ are gluons,
 $\gamma^\mu$ are the Dirac matrices and $g_s$ is the strong
interaction coupling constant.

     The $\lambda^a$ are the SU(3) color matrices, with
\beq
\label{lambda}
     \lambda^a \lambda^b-\lambda^b \lambda^a =i 2\sum_{c=1}^8 f^{abc} 
\lambda_c \; ,
\eeq
with $f^{abc}$ the SU(3) structure constants. See \cite{cl1984}
for a discussion of color, $\lambda^a$, $f^{abc}$, and quark-gluon interactions.

  Note that the strong interaction field, $A_\mu^a$, is similar to the familiar
electromagnetic (EM) field, $A^{EM}_\mu$, with the quantum of $A^{EM}_\mu$ 
being the photon. The strength of coupling of charged particles to photons is 
the electric charge $e$ for the eletromagnetic field, as $g$ is the strength of
coupling of quarks to gluons. The EM interaction of particles with electric
charge $e$ with photons is 
\beq
\label{epinteraction}
        V_{ep}&=& i\bar{\psi}\gamma^\mu e A^{EM}_\mu \psi
\; , 
\eeq
where $\psi$ is a quantum field with electric charge $e$.

    Since $g_s \gg e$, Feynman diagrams
can be used for the EM interaction but not for the strong interaction. 
Nonperturbative methods such as QCD sum rules used to derive the mixed heavy
quark hybrids, discussed below, must be used.

\subsection{Standard and hybrid heavy quark mesons}
In this review a main element is that the production of mixed heavy quark 
hybrid mesons can be used for the detection of the QGP via RHIC.

\newpage

In thus subsection we briefly discuss standard heavy quark and hybrid
heavy quark mesons. A standard meson consists of a quark and an antiquark with 
color = 0, while a hybrid meson consists of a quark and an antiquark with color
 = 8 bound with a gluon with color = 8, with the hybrid meson having color = 0. 
This is illustrated in the figures below, with the Heavy quark $Q=c$ 
(charmonium) or $Q=b$ (bottomonium).
\vspace{2cm}

\begin{figure}[ht]
\begin{center}
\epsfig{file=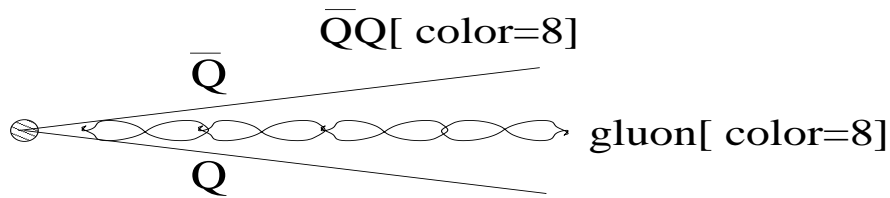,height=4cm,width=12cm}
\end{center}
\caption{A meson (color=0) formed by heavy quark bound with an anti-quark}
\end{figure}

\begin{figure}[ht]
\begin{center}
\epsfig{file=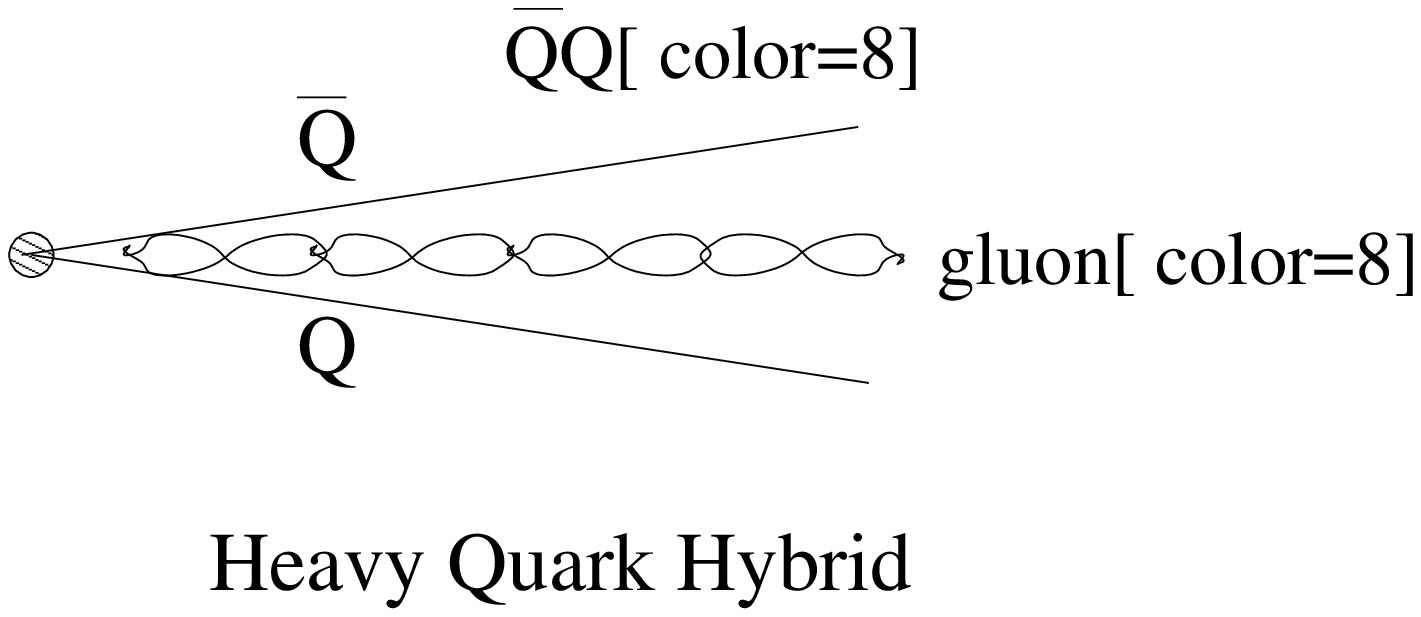,height=4cm,width=12cm}
\end{center}
\caption{A hybrid meson (color=0) formed by heavy quark anti-quark of color
8 bound with a gluon of color 8}
\end{figure}

\section{Classical and Quantum Phase Transitions}

A classical first order phase transition is the transformation of a system 
with a well-defined critical temperature, $T_c$  from one phase of matter to 
another. A quantum phase transition  is the transformation from  a quantum 
state to a different quantum state. The QCD Phase Transition (QCDPT) is a 
Cosmological phase transition in which the state of the universe transforms 
to a different state of the universe.
\newpage

\subsection{Classical Phase Transitions and Latent Heat}

  During a first order phase transition
as one adds heat the temperature stays at $T=T_c$ until all the matter has
changed to the new phase. The heat energy that is added is called latent
heat. 

 A familiar example of a first order phase transitions is ice, a solid, 
melting to form water, a liquid; and water boiling to form steam, a gas. 
Note that in addition to solid-liquid and liquid-gas, there are gas-plasma 
phase transitions, as is the QCDPT, discussed below.

\subsection{A Brief Review of Quantum Theory Needed for
Cosmological Phase Transitions}

 In quantum theory one does not deal with physical matter, but with states and
operators. A quantum phase transition is the transition from one state to a 
different state. For the study of Cosmological Phase Transitions a
state is the state of the universe at a particular time and temperature.

A quantum state represents the system, and a quantum 
operator operates on a state. For instance, a system is in state [1] and
there is an operator $A$.
\beq
        |[1]> &\equiv& {\rm state[1]} \nonumber \\
           A &\equiv& {\rm operator\;A} \; .
\eeq

  An operator operating on a quantum state produces another quantum state.
For example, operator $A$ operates on state [1]
\beq
      A|[1]> &=& |[2]> \; ,
\eeq
where state [2]=$|[2]>$ is a quantum state. If a system is in a quantum state, 
the value of an operator is given by the expectation valus. For example, 
consider state $|[1]>$ and operator $A$.
\beq
        <[1]|&\equiv& {\rm adjoint\;of\;state[1]} \nonumber \\
        <[1]|A|[1]> &\equiv& {\rm expectation\;value\;of\;A} \; .
\eeq

\subsection{Cosmological Phase Transitions}

  Calling $|0,T>$ the state of the universe at time t when it has temperature
$T$, an operator $A$ has the expectation value $<0,T|A|0,T>$, as discussed
above. If there is a cosmological first order phase transition, then there
is a critical temperature $T_c$ and
\beq
\label{DeltaA}
        <0,T|A|0,T>_{T<T_c}- <0,T|A|0,T>_{T>T_C}&=& \Delta A \; ,
\eeq
with $ \Delta A$ the latent heat of the cosmological phase transitions.

It has been shown via lattice gauge theory that the QCDPT is a first order
cosmological phase transformation that occured $t\simeq 10^{-5}$ seconds after 
the Big Bang, with the critical temperature  $kT^{QCDPT}_c\simeq 150 MeV$, where
$k$ is the Boltzman constant. For $t < 10^{-5}\;s$ the universe is a QGP,
whie  $t > 10^{-5}\;s$ the universe is our present universe.
\newpage

During the QCDPT bubbles of our present universe 
with protons, neutrons, etc (hadrons) nucleated within the universe with a 
dense plasma of quarks and gluons, the Quark-Gluon Plasma (QGP), that existed 
when the temperature of the universe was greater than $T^{QCDPT}_c$.

 We shall discuss the possible detection of the QGP via relativistic heavy 
ion collisions (RHIC),
but first we discuss the latent heat for the QCDPT and then the mixed hybrid
heavy quark states which might detect the QGP produced via RHIC.

\subsection{The QCDPT and Quark Condensate}

  As reviewed above, the QCD fermion fields and particles are quarks. The
Latent Heat for the QCD Phase Transition (QCDPT) is the Quark Condensate. 
With the notation $ q(x)$ = the quark field  and $\bar{q}(x)$ = the
antiquark field,
\beq
\label{quark}     
              |0,T > &=& {\rm\;vacuum\;state\;temperature=T} \nonumber \\
      <0,T|\bar{q}(x) q(x)|0,T> &=& {\rm \;quark\;condensate} \nonumber \\
       <0,T|\bar{q}(x) q(x)|0,T>&=& 0 {\rm \;in\;quark\;gluon\;plasma\;phase\;
T>T^{QCDPT}_c} \nonumber \\
           &\simeq& -(.23\;GeV)^3 {\rm \;in\;hadron\;phase\;T<T^{QCDPT}_c} 
\nonumber
\eeq

The quark condensate $<\bar{q}q>$ goes from 0 to $(.23)^3 GeV^3$ at the 
critical temperature $T^{QCDPT}_c$ of about 150 MeV, and is therefore a 
first order phase transition, with the latent heat = $(.23)^3 GeV^3$.
 
 Although we do not discuss Dark Energy in this review, note that
Dark Energy is cosmological vacuum energy, as is the quark condensate. It
has been shown that Dark Energy at the present time might have been created
during the QCDPT via the quark condensate\cite{zmk12}.

\section{Magnetic Wall From QCDPT and CMBR}

As a first order phase transition, during the QCDPT bubbles of our hadronic
universe form, collide, and form a magnetic wall. This wall produces B-type
Cosmic Microwave Background Radiation (CMBR) polarization correlations. 

\subsection{Magnetic wall produced during the QCDPT}

  The bubbles of the hadronic (nucleon) phase that form during the QCDPT, 
as shown in Fig.4, are composed mainly of glue. When two of the bubbles
collide they form an interior wall, similar to colliding soap bubbles, 
made of gluons. The nucleons within the hadronic bubble interact with the 
interior gluonic wall, and because of symmetry violation are oriented 
perpendicular to the wall. This creates a magnetic wall, shown in Figure 3
\newpage

\vspace{8.0cm}

\begin{figure}[ht]
\begin{center}
\epsfig{file=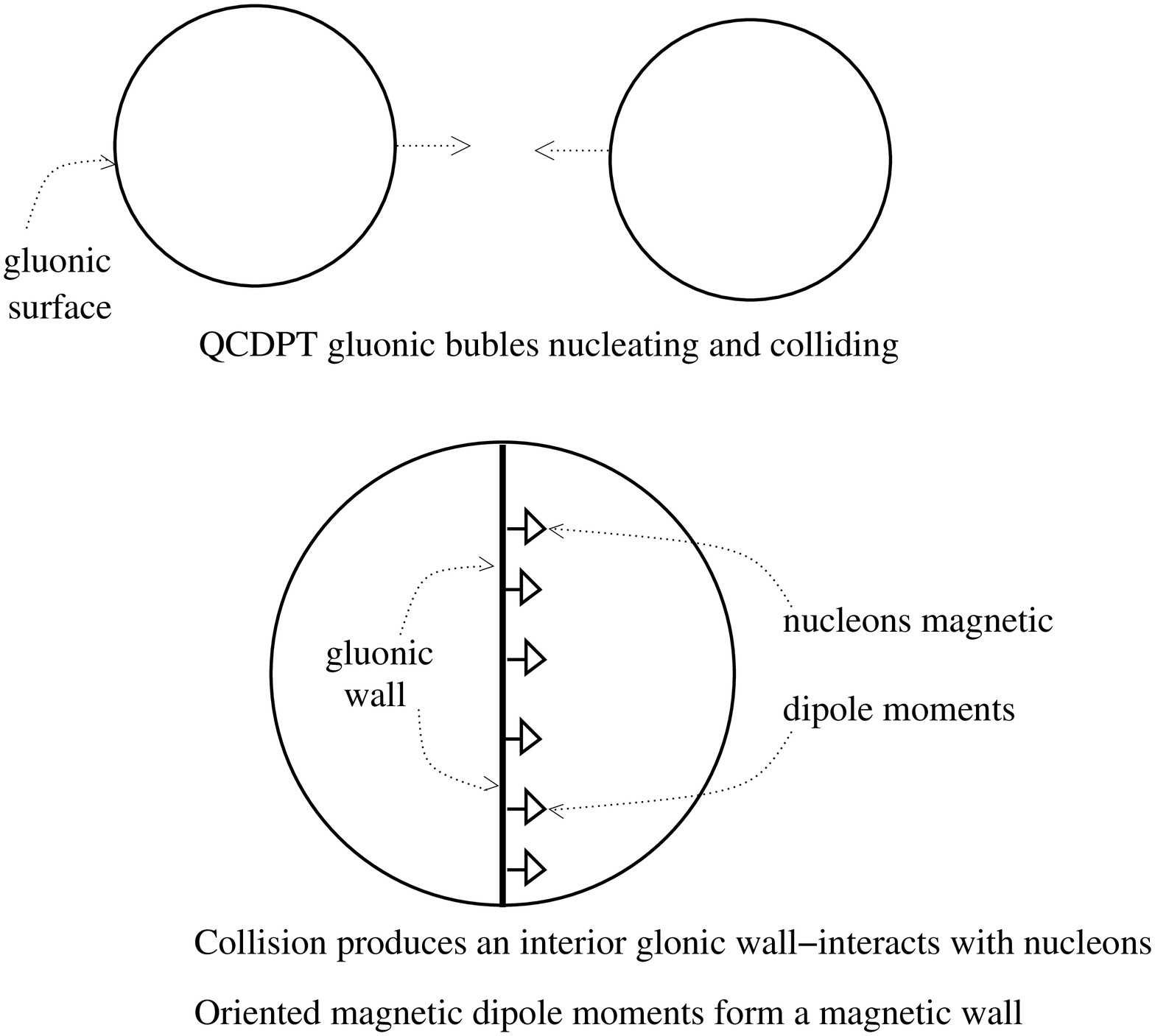,height=3cm,width=12cm}
\end{center}
\caption{Magnetic wall created during the QCDPT}
\end{figure} 

  The magnitude of the magnetic field (B-field) produced during the QCDPT
was estimated in \cite{lsk03} in an article motivated in part by improved 
CMBR observations, which promised polarization correlation measurements, and
that these magnetic structures could be primordial seeds of galactic and 
extra-galactic magnetic fields, a long-standing problem of astrophysics.

Using the mathematical form of \cite{fz2}, including estimates of the 
lifetime of the interior QCD gluonic wall, tnthe magnetic wall was
formed by the interaction of the nucleons with the gluonic wall. The
electromagnetic interaction (see Eq(\ref{epinteraction})) is
\beq
\label{VEMint}
     V^{EM}_{int} & = & i e \bar{\Psi} \gamma^\mu A^{EM}_\mu \Psi,
\eeq
where $\Psi$ is the nucleon field operator and $A^{EM}_\mu$ the EM field
discussed above.

This leads to the electromagnetic interaction with the nucleons magnetic
dipole moment given in terms of the electromagnetic field tensor, 
$F^{\mu\nu}$ by
\beq
\label{Vint}
      {\cal V}^{int} & = & \frac{e}{2 M_n} \bar{\Psi} \sigma_{\mu\nu}
\gamma_5 \Psi F^{\mu\nu},
\eeq
and a similar interaction (without the $\gamma_5$) for the electric dipole
moment, due to cp violation.  From Eq.(\ref{Vint}) one can estimate the 
magnetic field in the wall. For the gluonic instanton wall oriented in the 
x-y direction one obtains for $B_z \equiv B_W = F^{21}$ within the wall of 
thickness $\rho$
\beq
\label{bz}
     B_z & \simeq & \frac{1}{\rho\Lambda_{QCD}} \frac{e}{2 M_n} 
 < \bar{\Psi} \sigma_{21}\gamma_5 \Psi > .
\eeq
The matrix element in Eq.(\ref{bz}) is estimated using the
Fermi momentum in the plane of the wall during the QCD phase transition
as $\Lambda_{QCD}$, giving  $< \bar{\Psi} \sigma_{21}\gamma_5 \Psi > = 
\frac{4\pi}{(2\pi)^2} \Lambda_{QCD}^2$. The resulting magnitude of the magnetic 
field at the wall is (see \cite{hw97} for derivation of factor 3/14). 
\beq
\label{bw}
     B_W & \simeq & \frac{3 e}{14 \pi} \Lambda_{QCD}.
\eeq

Therefore this picture is that at the end of the QCD  phase
transition there is a magnetic wall in the hadronic phase in momentum space,
derived from the Fourier Transform of $B_W(x)$\cite{lsk03},
\beq
\label{wallk}
  {\bf B}_W({\bf k}) & = & \frac{B_W}{2\sqrt{2}b^2M_n} e^{-(k_x^2 + k_y^2)
/4b^2} e^{-k_z^2/4M_n^2},
\eeq
where $b^{-1}$ is of the scale of the horizon size, $d_H$, at the end of the 
QCDPT (t $\simeq 10^{-4}$ s), $b^{-1} =d_H \simeq$ few km, 
while $M_n^{-1} \simeq 0.2 fm$. 
\newpage

\subsection{CMBR Polarization Correlations From Magnetic Wall}

  From Einstein's General Theory of Relativity (see, e.g., Kolb-Turner,
''The Early Universe'') $T(t)$, the temperature of the universe at time t
is
\beq
\label{kT}
       kT(t)&\simeq& \frac{1 {\rm MeV}}{\sqrt{t({\rm in\;s})}}
\; .
\eeq

Using 380,000 years = $1.2 \times 10^{13}$,  1 MeV = $10^{6}$ s,

\beq
\label{kTcmb}
       kT(t=380,000 {\rm \;years}) &\simeq& 0.25 {\rm \;eV} \; .
\eeq  

Since the binding energy of electrons in hydrogen atoms is about 10 eV,
electrons bound to atoms and were no longer free to scatter EM waves from
the early universe. Light from the early universe was released: The CMBR.
\vspace{5mm}

 {\bf CMBR TEMPERATURE CORRELATIONS:}

  Aim two microwave telescopes seperated by the angle $\theta$ into the sky

\begin{figure}[ht]
\begin{center}
\epsfig{file=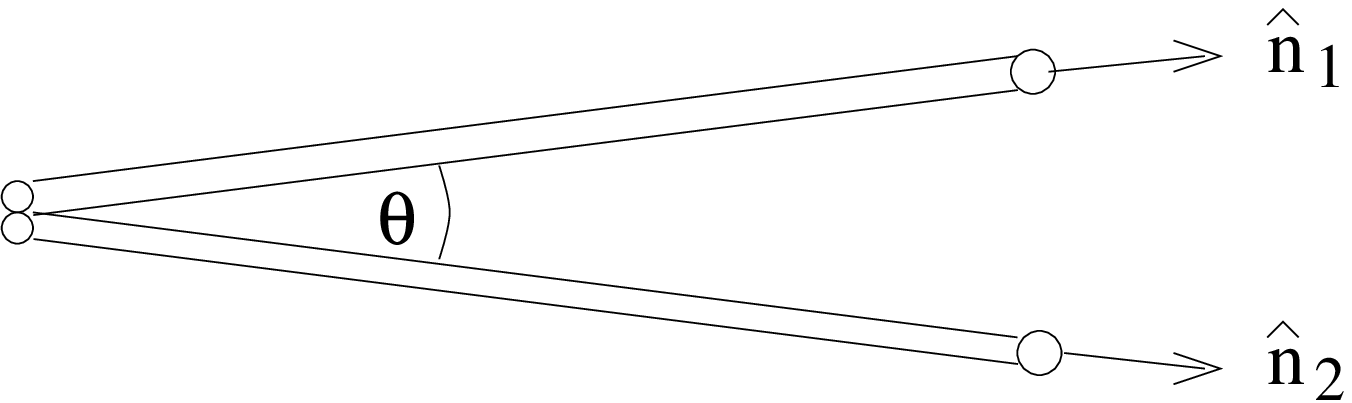,height=4cm,width=12cm}
\caption{}
\end{center}
\end{figure}

Expand the temperature differences, $ \Delta T(\hat{n})= T(\hat{n})-T_o$ in 
terms of angular functions
\beq
\label{CTT}
    <\frac{\Delta T(\hat{n}_1)}{T_0}\frac{\Delta T(\hat{n}_2)}{T_0}> &=& 
 \sum_{lm}C^{TT}_l Y_{lm}(\hat{n}_1)Y_{lm}(\hat{n}_2)
\nonumber \\
 {\rm use}\sum_{m}Y_{lm}(\hat{n}_1)Y_{lm}(\hat{n}_2) &=& \frac{2 l +1}{4\pi}
P_l(cos\theta) \nonumber \\
 {\rm giving} <\frac{\Delta T(\hat{n}_1)}{T_0}\frac{\Delta T(\hat{n}_2)}{T_0}>
 &=&  \sum_{l}\frac{2 l +1}{4\pi} C^{TT}_l P_l(cos\theta) 
\eeq

 The $P_l$ are well-known Legendre polynomials. 
The $C^{TT}_l$ are the measured Temperature-Temperature correlations.
From astrophysics observations COBE\cite{cobe96} WMAP\cite{wmap13} 
ACBAR\cite{acbar08} QUaD\cite{quad10} the following was found:

\hspace{3cm} The universe is flat (space is not curved)

\hspace{3cm} Baryon density = 0.04 (4 $\%$ of density of university)

\hspace{3cm} Dark Matter density = 0.23 (what is dark matter?)

\hspace{3cm} Dark energy density = 0.73 (vacuum energy)

\newpage
 {\bf CMBR B-B CORRELATIONS:}

   B-type polarization correlations, $C^{BB}_l$ are derived from 
$<B_z({\bf k},\eta)B_z({\bf k'},\eta)>$, with ${\bf k} = k\hat{n}$ and
$\eta$ is conformal time,  similar to $C^{TT}_l$ being
derived from  $ <\Delta T(\hat{n}_1) \Delta T(\hat{n}_2>$ (Eq(\ref{CTT})).

To get the power spectrum one must evaluate 
$<B_z({\bf k},\eta)B_z({\bf k'},\eta)>$. Using the fact that
$exp(-k^2 d_H^2) \simeq 1.0$ at the time of the QCDPT,
\beq
\label{bcor}
   <B_z({\bf k},\eta)B_z({\bf k'},\eta)> & \simeq & {\cal B}_W^2
 \delta(k_x-k'_x) \delta(k_y-k'_y)<e^{-k_z^2/4M_n^2}e^{-k^{'2}_z/4M_n^2}> 
\nonumber \\
       & \simeq & {\cal B}_W^2 d_H e^{-k_z^2/4M_n^2} \delta({\bf k-k'}).
\eeq 
This gives for the polarization power spectrum (see \cite{hw97})
\beq
\label{Bpower}
   C^{BB}_l & = & \frac{(l+1)(l+2)}{\pi} {\cal B}_W^2 d_H \int dk 
 \frac{j_l^2[k(\Delta \eta)]}{(\Delta \eta)^2},
\eeq
where the conformal time integral has been carried out and $\Delta \eta$ is 
the conformal time width at the last scattering. The integral over the 
spherical Bessel function is carried out by using the fact that 
$\int dz j_l^2(z) = \pi/(4l)$ for large $l$. Therefore in the range
$ 100 < l < 2000$ one has the approximate result
\beq
\label{cbb}
    C^{BB}_l & \simeq & \frac{25 d_H^5 B_W^2}{1152 M_n^2 \Delta\eta^3}l^2.
\eeq
Using the parameters $M_n\Delta\eta =1.5 \times 10^{39}$ \cite{ss}, $d_H = 
0.37\times 10^{24} GeV^{-1}$, and $B_W = 1.0\times10^{17}$ Gauss, 
\beq
\label{cbbf}
    C^{BB}_l & \simeq & 4.25\times10^{-8} l^2
\eeq

The result for the B-type power spectrum is shown in Fig. 9 by the
solid line.
\vspace{-2cm}

\begin{figure}[ht]
\begin{center}
\epsfig{file=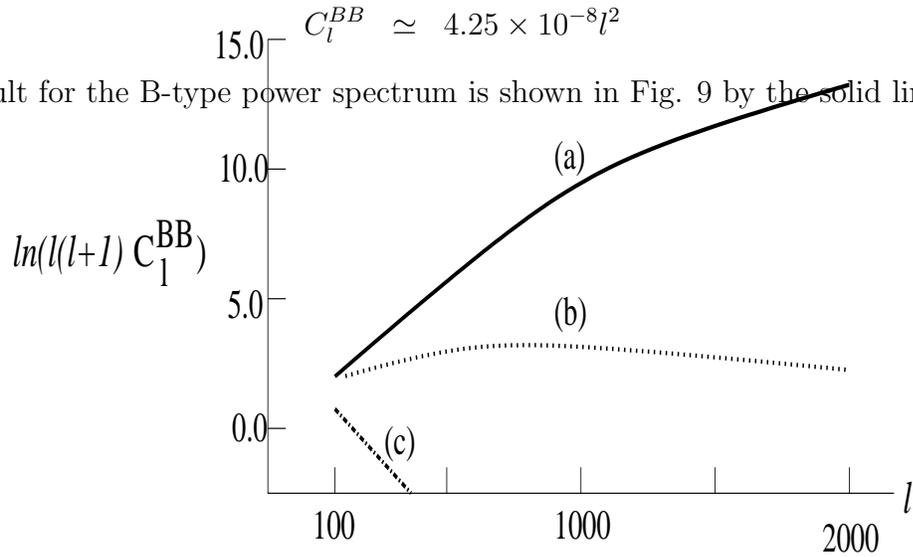,height=7cm,width=12cm}
\caption{B power spectrum in (a) magnetic wall theory, (b) string theory,
 (c) inflation theory}
\end{center}
\end{figure}

  The magnetic wall prediction for $C^{BB}_l$ differs from other cosmological 
predictions, which should be measurable with the next generation of CMBR 
measurements.
\newpage
\section{Mixed Hybrid Heavy Quark Meson States}

  The Charmonium $\psi'(2S)$ and Upsilon $\Upsilon(3S)$ states, which are 
important for this review, are not consistent with being standard $c \bar{c}$ 
and $b \bar{b}$ mesons as has been shown by their decays into hadrons.

\subsection{Heavy quark meson decay puzzles} 

A puzzle for Charmonium meson states is the ratio of branching ratios for 
$\psi'(2S)$ vs $J/\Psi(1S)$ to decay to  hadrons. For example the
$\rho-\pi$ puzzle: The $\Psi'(2S)$ to $J/\Psi$ ratios for $\rho-\pi$ 
and decays are more than an order of magnitude smaller than predicted
by the standard model\cite{frannk83,kpr08}.

A puzzle for upsilon meson states is the two $\pi$ decay, with the 
$\sigma$ a broad 600 MeV $\pi-\pi$ resonance is:
\vspace{5mm}

$\Upsilon(2S) \rightarrow \Upsilon(1S) + 2\pi$ large branching ratio. 
No $\sigma$ 
\vspace{5mm}

$\Upsilon(3S) \rightarrow \Upsilon(1S) + 2\pi$ large branching ratio to 
$\sigma$ 
\vspace{5mm} 

This has been called Vogel $\Delta n=2$ Rule\cite{lsk09,vogel}.
Neither of these puzzles can be solved using standard QCD models. They were
solved using the mixed heavy hybrid theory.

\subsection{Mixed charmonium-Hybrid charmonium States}

 The method of QCD Sum Rules\cite{sz79} was used to study the heavy quark
Charmonium and Upsilon states, and show that two of them are mixed hybrid
meson states\cite{lsk09}

 The starting point of the method of QCD sum rules\cite{sz79} for finding 
the mass of a state A is the correlator,
 
\beq
\label{2PiA(x)}
       \Pi^A(x) &=&  \langle | T[J_A(x) J_A(0)]|\rangle \; ,
\eeq
with $| \rangle$ the vacuum state and
the current $J_A(x)$ creating the states with quantum numbers A.
The following mixed vector ($J^{PC}=1^{--}$) charmonium, hybrid charmonium 
current was used in QCD Sum Rules\cite{lsk09}
\beq
\label{11}
        J^\mu &=& b J_H^\mu + \sqrt{1-b^2} J_{HH}^\mu 
\eeq
with
\beq
\label{12}
          J_H^\mu &=& \bar{q}_c^a \gamma^\mu q_c^a \nonumber \\
          J^\mu_{HH} &=&  \bar{\Psi}\Gamma_\nu G^{\mu\nu} \Psi \; ,
\eeq 
where $\Psi$ is the heavy quark field, $\Gamma_\nu = C \gamma_\nu$,
$\gamma_\nu$ is the usual Dirac matrix, C is the charge conjugation operator,
and the gluon color field is
\beq
\label{Gmunu}
         G^{\mu\nu}&=& \sum_{a=1}^8 \frac{\lambda_a}{2} G_a^{\mu\nu}
\; ,
\eeq
\newpage

with $\lambda_a$ the SU(3) generator ($Tr[\lambda_a \lambda_b]= 2 \delta_{ab}$), 
discussed above.

 Using QCD sumrules it was found\cite{lsk09} that
the  $\Psi'(2S)$ state is 50\% normal and 50\% hybrid. The
analysis for upsilon states was similar, with the $\Upsilon(3S)$ being 50\% 
normal and 50\% hybrid: 
\beq
 |\Psi'(2s)>&=& -0.7 |c\bar{c}(2S)>+\sqrt{1-0.5}|c\bar{c}g(2S)> \\
 |\Upsilon(3S)>&=& -0.7 |b\bar{b}(3S)>+\sqrt{1-0.5}|b\bar{b}g(3S)> \; 
\eeq. 

The QCD sum rule method is nonperturbative, has proved to be accurate in 
many calculations, with a 10\% error in $b$, Eq(\ref{11}), and $M_B^2$.

\section{Heavy Quark State Production In p-p and A-A Collisions}

In order to discuss the possible detection of a QGP using the production of
mixed heavy quark hyprids in A-A collisions (RHIC), we first review the
earlier research on  production of mixed heavy quark hyprids in p-p collisions,
and then recent research on A-A collisions for E=$\sqrt{s}$= 200 GeV.

\subsection{Production of $\Psi$ and $\Upsilon$ States In p-p Collisions}

In this subsection we review the publication of \cite{klm11} on heavy 
quark state production in unpolarized p-p collisions.

The color octet model\cite{cl96,bc96,fl96,cl296} was used.  The three octet 
matrix elements needed are $<O_8^\Phi(^1S_0)>$, $<O_8^\Phi(^3S_1)>$, and 
$<O_8^\Phi(^3P_0)>$, with $\Phi$ either $J/\Psi$, $\Psi'(2S)$, or $\Upsilon(nS)$.
In \cite{klm11} the second scenario of Nyyak and Smith\cite{ns06} was used:
\beq
\label{octetmatrixelements}
 <O_8^\Phi(^1S_0)>&=&.039 {\rm \;} <O_8^\Phi(^3S_1)>=.0112 {\rm \;and\;} 
<O_8^\Phi(^3P_0)> =0 \; .
\eeq

The differential production cross sections for $\Phi$ for helicity 
$\lambda$ = 0 and 1 need the parameter $A_\Phi$ with
\beq
\label{Aphi}
       A_\Phi&=&\frac{5 \pi^3 \alpha_s^2}{288 m^3 s}<O_8^\Phi(^1S_0)>
\;,
\eeq
with $s=E^2$, the strong coupling constant $\alpha_s \simeq 0.119$, and 
$m=$ 1.5 or 5 GeV for the charmonium or bottomonium quark mass.

  For the  differential cross sections the rapidity variable, $y$, is used.
The differential rapidity distributions for $\lambda=0$,  $\lambda=1$ are 
given by (with $a= 4m^2/s$)
\beq
\label{5}
      \frac{d \sigma_{pp\rightarrow \Phi(\lambda=0)}}{dy} &=& 
     A_\Phi \frac{1}{x(y)} f_g(x(y),2m)f_g(a/x(y),2m) \frac{dx}{dy} \; ,
\eeq
\beq
\label{6}
\frac{d \sigma_{pp\rightarrow \Phi(\lambda=1)}}{dy} &=& A_\Phi \frac{1}{x(y)}
[f_g(x(y),2m)f_g(a/x(y),2m)+0.613(f_d(x(y),2m)f_{\bar{d}}(a/x(y),2m) 
\nonumber \\
    &&+f_u(x(y),2m)f_{\bar{u}}(a/x(y),2m)]\frac{dx}{dy} \; ,
\eeq
\newpage

with $f_g(x,2m)$, $f_q(x,2m)$ the gluonic and quark distribution functions 
evaluated at $Q=2m$. See \cite{klm11} for derivation of the factor 0.613. 
The variable $x(y)$ 
is
 \beq
\label{x(y)}
      x(y) &=& 0.5 \left[\frac{m}{\sqrt{s}}(\exp{y}-\exp{(-y)})+
\sqrt{(\frac{m}{\sqrt{s}}(\exp{y}-\exp{(-y)}))^2 +4a}\right] \; .
\eeq

Using CTEQ6\cite{CTEQ6} for $Q=2m_c$ = 3 GeV and  $\sqrt{s} \simeq$ 200 GeV,
and for $Q=2m_b$ = 10 GeV and $\sqrt{s} \simeq$ 38.8 GeV
 $f_g(x,2m)$ and $f_q(x,2m)$ were obtained. 

  Note that $A_\Phi$ is enhanced by a factor $\simeq \pi$ for the mixed hybrid
$|\Psi'(2s)$ and $|\Upsilon(3S)>$ states as derived in \cite{klm11}.

 \subsection{Charmonium  Production Via Unpolarized p-p
Collisions at E=$\sqrt{s}$= 200 Gev}

For unpolarized p-p collisions with $\sqrt{s}=200 GeV$ using 
scenario 2\cite{ns06}, with the 
nonperturbative matrix elements given above, $A_\Phi=
\frac{5 \pi^3 \alpha_s^2}{288 m^3 s} <O_8^\Phi(^1S_0)>$ =$7.9 \times 10^{-4}$nb 
for $\Phi$=$J/\Psi$; $a= 4m^2/s = 2.25 \times 10^{-4}$ for Charmonium.
For  $\sqrt{s}=200 GeV$ $x(y)$ is obtained rfom Eq(\ref{x(y)}) and
\beq
\label{7}
  \frac{d x(y)}{d y} &=&\frac{M}{400}(\exp{y}+\exp{(-y)})\left[1. + 
\frac{\frac{M}{200}(\exp{y}+\exp{(-y)})}{\sqrt{(\frac{M}{200} 
(\exp{y}-\exp{(-y)}))^2 +4a}}\right] \; .
\eeq
\vspace{4cm}

\begin{figure}[ht]
\begin{center}
\epsfig{file=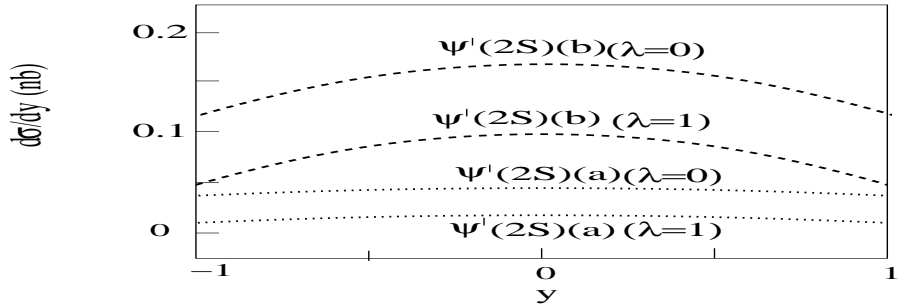,height=4cm,width=12cm}
\caption{d$\sigma$/dy for Q= 3 GeV, E=200 GeV unpolarized p-p  collisions 
producing$\Psi'(2S)$  with $\lambda=1$,$\lambda=0$. Reults labeled 
$\Psi'(2S)(a)$ are  obtained using the standard model; results labeled
 $\Psi'(2S)(b)$ are obtained by using the mixed hybrid theory }
\label{Figure 5}
\end{center}
\end{figure}
 
\newpage

  Using Eqs(\ref{5},\ref{6},\ref{7}), with the parton distribution functions 
given in Ref\cite{klm11}, we find d$\sigma$/dy for Q=3 GeV, $\lambda=0$ and 
$\lambda=1$ the results for $\Psi'(2S)$  shown in Figure 6.
The results for d$\sigma$/dy shown in Figure 6 labeled $\Psi'(2S)(a)$ are
 obtained by using for the standard nonperturbative matrix element=0.039 
times the matrix elements for $J/\Psi$ production\cite{klm11}, while the 
results labeled  $\Psi'(2S)(b)$ are obtained by using the mixed hybrid matrix e
lement enhanced by a factor of $\pi$.

\subsection{Upsilon Production Via Unpolarized p-p
Collisions at E=$\sqrt{s}$= 38.8 Gev}

For Q=10 GeV, using the parton distributions given in Ref\cite{klm11} and 
Eqs(\ref{5},\ref{6}) for helicity $\lambda=0$, $\lambda=1$, with
$A_{\Upsilon}$ =$5.66 \times 10^{-4}$nb and $a=6.64 \times 10^{-2}$, one obtains 
$d\sigma/dy$ for $\Upsilon(nS)$ production, $d\sigma/dy$  for  
$\Upsilon(3S)$ are shown in Figure 7.
\vspace{6cm}

\begin{figure}[ht]
\begin{center}
\epsfig{file=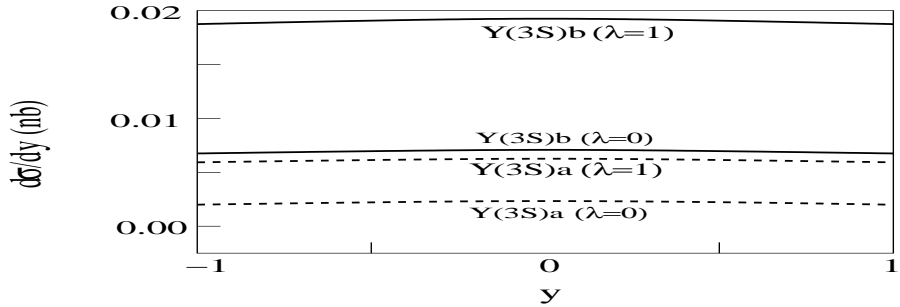,height=4cm,width=12cm}
\caption{d$\sigma$/dy for Q= 10 GeV, E=38.8 GeV unpolarized p-p collisions 
producing $\Upsilon(3S)$ with $\lambda=0$, $\lambda=1$. Theresults for the 
standard model are labelled $\Upsilon(3S)$a, while for the hybrid theory are 
labelled $\Upsilon(3S)$b}
\label{Figure 10}
\end{center}
\end{figure}

In Fig. 7 as with $\Psi'(2S)$, the enhancement for $\Upsilon(3S)$
is a factor of $\pi$ for the mixed hybrid theory vs. the standard
model.

 It should be noted that the ratios of $d\sigma/dy$ for $\Psi'(2S)$,
 and $\Upsilon(3S)$, for the hybrid theory vs. 
the standard are our most significant results. As will be discussed below,
this might be used for detection of the QGP via  relativistic energy A-A
collisions (RHIC).
\clearpage
\subsection{Heavy-quark state production in A-A collisions
 at $\sqrt{s_{pp}}$=200 GeV}

 This subsection is a review of \cite{klm14}, which is an extension of
the theory of \cite{klm11} for p-p to A-A collisions.

 The differential rapidity cross section for the production of a heavy
quark state with helicity $\lambda=0$ in the color octet model in A-A
collisions is given by

\beq
\label{1}
   \frac{d \sigma_{AA\rightarrow \Phi(\lambda=0)}}{dy} &=& 
   R^E_{AA} N^{AA}_{bin}< \frac{d \sigma_{pp\rightarrow \Phi(\lambda=0)}}{dy}>
\; ,
\eeq
where $R^E_{AA}$ is the product of the nuclear modification factor $R_{AA}$
and $S_{\Phi}$, the dissociation factor after the state $\Phi$ (a charmonium or
bottomonium state) is formed (see \cite{star09}). $N^{AA}_{bin}$ is the number
of binary collisions in the AA collision, and 
$< \frac{d \sigma_{pp\rightarrow \Phi(\lambda=0)}}{dy}>$ is the
differential rapidity cross section for $\Phi$ production via nucleon-nucleon
collisions in the nuclear medium. Note that $R^E_{AA}$, which
we take as a constant, can be functions of rapidity\cite{vogt08,vitov12}.

  Experimental studies show that for $\sqrt{s_{pp}}$ = 200 GeV 
$R^E_{AA}\simeq 0.5$ both for Cu-Cu\cite{star09,phenix08} and 
Au-Au\cite{phenix07,star07,kks06}. The
number of binary collisions are  $N^{AA}_{bin}$=51.5 for Cu-Cu\cite{sbstar07} 
and 258 for Au-Au. The differential rapidity cross section for pp collisions 
in terms of $f_g$, the gluon distribution function
($-0.8\leq y \leq 0.8$ for $\sqrt{s_{pp}}$ = 200 GeV with $f_g$ from 
\cite{klm11}), is

\beq
\label{2}
     < \frac{d \sigma_{pp\rightarrow \Phi(\lambda=0)}}{dy}> &=& 
     A_\Phi \frac{1}{x(y)} f_g(\bar{x}(y),2m)f_g(a/\bar{x}(y),2m) 
\frac{dx}{dy} \; ,
\eeq  
     where $a= 4m^2/s$; with $m=1.5$  GeV for charmonium, and 5 GeV for 
bottomonium. For $\sqrt{s_{pp}}$ = 200 GeV $A_\Phi=7.9 \times 10^{-4}$nb for 
$\Phi$=$J/\Psi$ and $2.13 \times 10^{-5}$nb for $\Upsilon(1S)$; $a = 
2.25 \times 10^{-4}$ for Charmonium and $2.5 \times 10^{-3}$ for Bottomium.

 The function $\bar{x}$, the effective parton x in a nucleus (A), is given in  
\cite{vitov06,vitov09}:
\beq
\label{barx}
         \bar{x}(y)&=& x(y)(1+\frac{\xi_g^2(A^{1/3}-1)}{Q^2}) \nonumber \\
   x(y) &=& 0.5 \left[\frac{m}{\sqrt{s_{pp}}}(\exp{y}-\exp{(-y)})+
\sqrt{(\frac{m}{\sqrt{s_{pp}}}(\exp{y}-\exp{(-y)}))^2 +4a}\right] \;,
\eeq
with\cite{qiu04} $\xi_g^2=.12 GeV^2$. For $J/\Psi$  $Q^2=10 GeV^2$, so
$\bar{x}=1.058 x$ for Au and $\bar{x}=1.036 x$ for Cu, while for $\Upsilon(1S)$
$Q^2=100 GeV^2$, so $\bar{x}=1.006 x$ for Au and $\bar{x}=1.004 x$ for Cu.

From Eqs(\ref{1},\ref{2},\ref{barx}) and the parameters given above
one finds the differential rapidity cross sections for  $J/\Psi$, $\Psi(2S)$,
and $\Upsilon(nS)$ states for A-A collisions.

The differential rapidity cross sections for $\Psi(2S)$ and $\Upsilon(3S)$ 
production via Cu-Cu and Au-Au collisions at RHIC (E=200 GeV) are shown in the 
figures below for both the standard model and the mixed hybrid theory
with the cross sections enhanced by $~\pi$ with the mixed hybrid theory,
 as discussed above.

 The absolute magnitudes are uncertain, and the shapes and relative magnitudes 
are the main prediction.

\clearpage

 The differential rapidity cross sections are shown in the 
following figures for $\Psi(2S)$ and $\Upsilon(3S)$ mixed hybrid statre
 production via Cu-Cu, Au-Au collisions (E=200 GeV).
\vspace{1.5cm}

\begin{figure}[ht]
\begin{center}
\epsfig{file=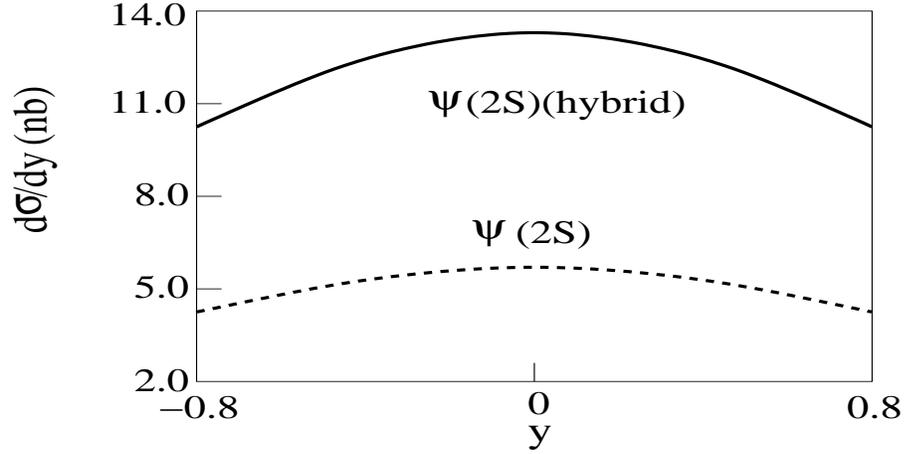,height=6 cm,width=12cm}
\caption{d$\sigma$/dy for 2m=3 GeV, E=200 GeV Cu-Cu collisions 
producing $\Psi(2S)$ with $\lambda=0$. The dashed curve is for the standard
$c\bar{c}$ model.}
\label{Figure 5}
\end{center}
\end{figure}
\vspace{2cm}

\begin{figure}[ht]
\begin{center}
\epsfig{file=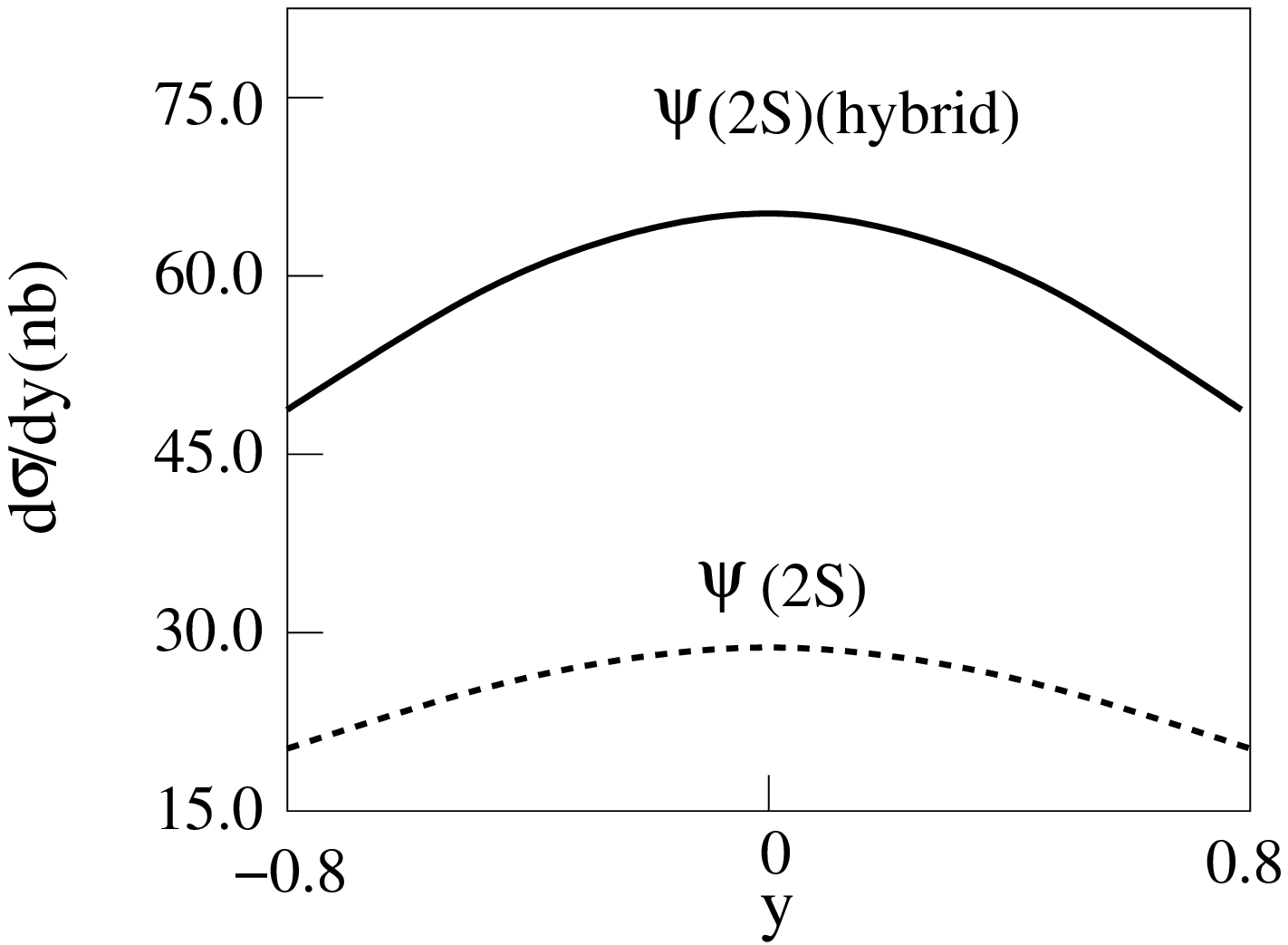,height=6 cm,width=12cm}
\caption{d$\sigma$/dy for 2m=3 GeV, E=200 GeV Au-Au collisions 
producing $\Psi(2S)$ with $\lambda=0$. The dashed curve is for the standard
$c\bar{c}$ model.}
\label{Figure 6}
\end{center}
\end{figure}

\clearpage

\begin{figure}[ht]
\begin{center}
\epsfig{file=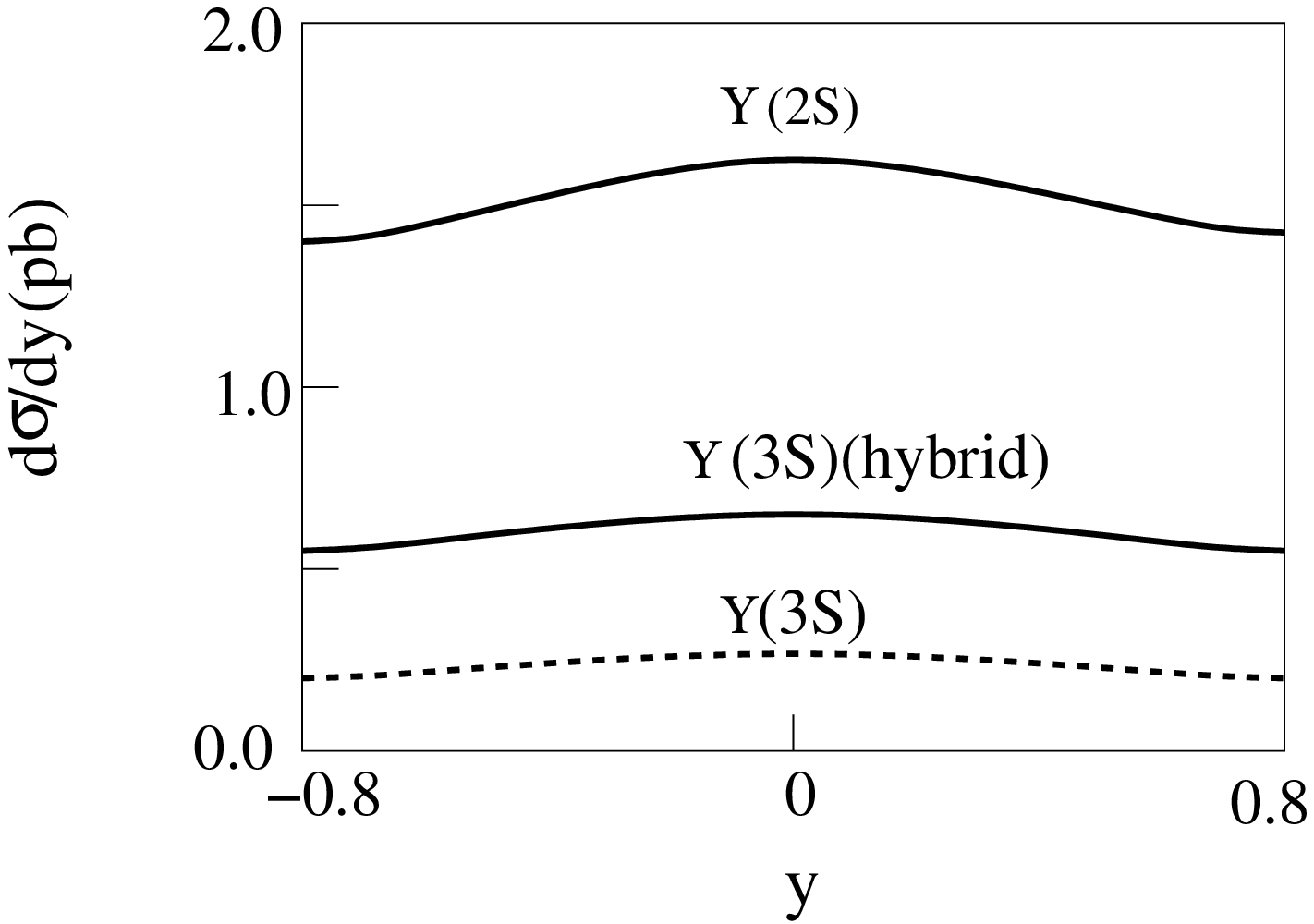,height=7 cm,width=12cm}
\caption{d$\sigma$/dy for 2m=10 GeV, E=200 GeV Cu-Cu collisions 
producing $\Upsilon(2S),\Upsilon(3S)$ with $\lambda=0$. For $\Upsilon(3S)$
the dashed curve is for the standard $b\bar{b}$ model.}
\label{Figure 9}
\end{center}
\end{figure}
\vspace{5 cm}

\begin{figure}[ht]
\begin{center}
\epsfig{file=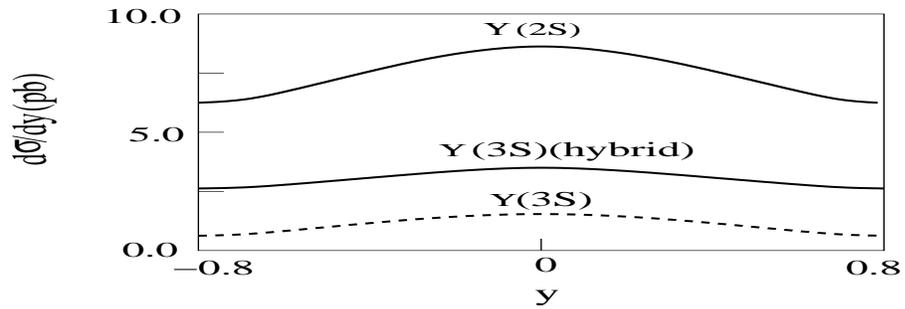,height=4 cm,width=12cm}
\caption{d$\sigma$/dy for 2m=10 GeV, E=200 GeV Au-Au collisions 
producing $\Upsilon(2S),\Upsilon(3S)$ with $\lambda=0$. For $\Upsilon(3S)$
the dashed curve is for the standard $b\bar{b}$ model.}
\label{Figure 10}
\end{center}
\end{figure}
\newpage

\subsection{Creation  of the QGP}

  At Brookhaven National Lab (BNL) the energy has been increaased so that
with RHIC, such as Pb-Pb collisions,  the energy of the atomic nuclei is
large enough so just after the nuclei collide the temperature is
that of the unverse about $10^{-5}$ seconds after the Big Bang. That is
$k T_c^{QCDPT} \simeq$ 150 MeV. This results in
the reverse QCDPT during which the nucleons, consisting of three quarks bound
by interactions with gluons, transform to the QGP, a dense plasma of quarks and
gluons. This is illustrated in the figure below.

\begin{figure}[ht]
\begin{center}
\epsfig{file=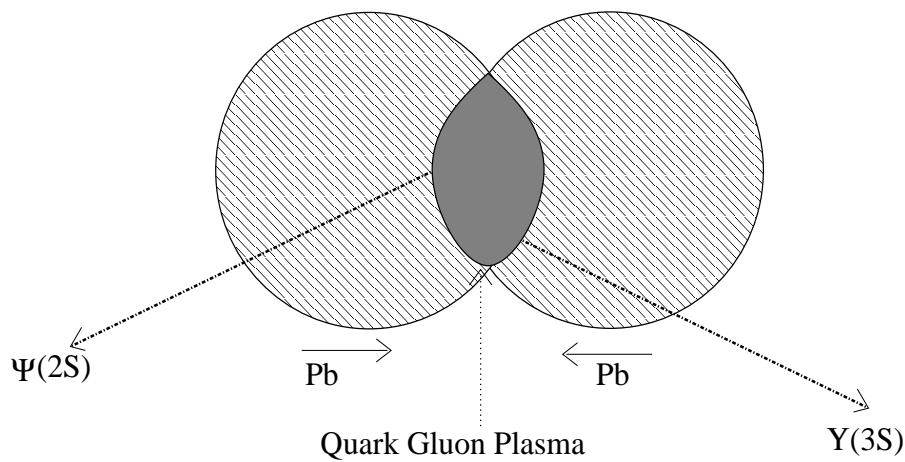,height=6 cm,width=12cm}
\caption{ Pb-Pb collisions producing the QGP, which produces $\Psi(2S)$ and 
$\Upsilon(3S)$ states.}
\end{center}
\end{figure}
\subsection{Detection  of the QGP via Hybrid Heavy Quark State 
Production}

  Because of the active gluons in the hybrid components of $\Psi(2S)$ and
$\Upsilon(3S)$ heavy quark mesons, the production of these mixed hybrid
states by the QGP, which also has active gluons, is a possible test of the
creation of the QGP via RHIC. 

This would be a remarkble acheivement for
nuclear and particle physics. It would also be important for cosmology,
in which it is believed that the universe before $10^{-5}$ seconds after the 
Big Bang consists of the QGP, but no experimental evidence of the QCD
exists at the present time. 

  In summary, in future experiments at BNL and the LHC-CERN the study of RHIC 
producing $\Psi(2S)$ and $\Upsilon(3S)$ mixed hybrid mesons could be a 
method for determining the creation of the QGP via atomic nuclei colliding
with an energy large enough to create matter with a temperature $T_c^{QCDPT}$,
the critical temperature for the cosmoligical phase transition from the
QGP to our present universe with nucleons.

\newpage
\subsection{Ratios of $\Psi'(2S)$ to $J/\Psi$ and $\Upsilon(2S)$,
 $\Upsilon(3S)$  to $\Upsilon(1S)$ cross sections cross sections}

Since the cross section ratios of the standard and mixed hybrid $\Psi$
and $\Upsilon$ states were discussed in detail in Ref\cite{kd16},
we shall not review these ratios.

\section{$\Psi(2S)$, $\Upsilon(3S)$ Suppression in p-Pb, Pb-Pb 
Collisions and Mixed Hybrid Theory}

 The suppressin of $\Psi$ and $\Upsilon$ states in passing through a nucleas
is discussed in Ref\cite{ks95,lsk16,kdz16}

The probability for production of a heavy quark meson state $\Phi$ on a nucleus,
 with $A$ the number of nucleons, irelative to a nucleon, is the suppression 
$S_A$ with
\beq
\label{SA}
       S_A &=& e^{-n_o\sigma_{\Phi N} L} \; ,
\eeq
where $L$ is the length of the path of $\Phi$ in nuclear matter $\simeq$ 10 fm 
for p-Pb collisions and 15 fm for Pb-Pb collisions, $\sigma_{\Phi N}$ is the 
cross section for $\Phi$-nucleon collisions, and the nuclear matter density 
$n_o\simeq .07 fm^{-3}$. Note that in Refs\cite{lsk16,kdz16} there were typos
sith $n_o=.017 fm^{-3}$.

\subsection{$\Psi(2S)$ to  $J/\Psi(1S)$ suppression in p-Pb
collisions}

 The cross section for standard charmonium $c\bar{c}$ meson via strong
QCD interactions with nucleons is given by\cite{ks95}
\beq
\label{sigma-c-barc} 
   \sigma_{c \bar{c} N} &=& 2.4 \alpha_s \pi r_{c \bar{c}}^2 \; ,
\eeq
where the strong coupling constant $\alpha_s\simeq $ 0.118, and 
the charmonium meson radius $r_{c \bar{c}} \simeq \not h/(2 M_c c)$, with $M_c$ 
the charm quark mass. Using $2 M_c \simeq M_{J/\Psi} \simeq $ 3 GeV,
$ r_{c \bar{c}}  \simeq  \not h/(3 GeV c) \simeq 6 \times 10^{-17}m = 0.06 fm$,
giving
\beq
\label{sigmacbarc}
        \sigma_{c \bar{c} N} &\simeq & 3.2 \times 10^{-3} fm^2=
3.2 \times 10^{-2} mb \; .
\eeq

Therefore $ n_o\sigma_{c \bar{c} N} L \simeq 0.0022$ and
\beq
\label{SAcc}
          S_A^{c\bar{c}} &=& e^{- n_o\sigma_{c \bar{c} N} L} \simeq 1.0 \; .
\eeq

 The cross section for hybrid charmonium $c\bar{c}g$ meson 
via strong QCD interactions with nucleons has been estimated in Ref\cite{ks95}
as $\sigma_{c\bar{c}g N} \simeq$ 6.5 mb. Therefore, $n_o\sigma_{c\bar{c}g N} L
\simeq 1.1$, and
\beq
\label{SAccg}
   S_A^{c\bar{c}g} &\simeq&  0.33  \; .
\eeq 

Using the mixed hybrid  theory for $\Psi(2S)$, with 50\% $|c\bar{c}>$ and 
50\% $|c\bar{c}g>$, the relatrive suppression of $\Psi(2S)$ to $J/\Psi(1S)$
is
\beq
\label{SAratio}
    R^{\Psi(2S)-J/\Psi(1S)}|_{theory}&\simeq& \frac{1+ 0.33}{2} 
\nonumber \\
          &=& 0.66 \; .
\eeq 

The experimental result\cite{ALICE14} is
\beq
\label{exp-suppression}
        R^{\Psi(2S)-J/\Psi(1S)}|_{exp}&\simeq& 0.65 \pm 0.1 \; ,
\eeq  
therefore the theory agrees with the experiment within errors.

\subsection{Suppression of  $\Upsilon(3S)$/$\Upsilon(1S)$ in Pb-Pb 
Collisions}

  The experimental results for the ratio of ratios needed for the present
subsection is\cite{cms12} is
\beq
\label{3S1SPbpp}
  \frac{\Upsilon(3S)/\Upsilon(1S)|_{PbPb}}{\Upsilon(3S)/\Upsilon(1S)|_{pp}}
     &=& 0.06 \pm 0.06 ({\rm stat})\pm 0.06 ({\rm syst}) ; .
\eeq

Using  $\sigma_{b\bar{b} N}\simeq \sigma_{c\bar{c} N}(M_c/M_b)^2\simeq 0.09 
\sigma_{c\bar{c} N}$, one obtains for the $b\bar{b}$ component of $\Upsilon(3S)$
\beq
\label{upsilonstandard}
 S_A^{b\bar{b}}&=&\frac{\Upsilon(3S)/\Upsilon(1S)|_{PbPb}}
{\Upsilon(3S)/\Upsilon(1S)|_{pp}}|_{sm} \simeq 0.11 \; .
\eeq

 For the  hybrid bottomonium $b\bar{b}g$ meson $\sigma_{b\bar{b}g N}=
\sigma_{c\bar{c}g N}(M_c/M_b)^2\simeq 0.09 \sigma_{c\bar{c}g N} \simeq 0.059 fm^2$,
and with $L \simeq$ 15 fm for Pb-Pb collisions, $n_o\sigma_{b\bar{b}g N} L \simeq
 0.15$, giving $S_A^{b\bar{b}g} \simeq 0.017$. From this and 
Eq(\ref{upsilonstandard}) one finds\cite{lsk16}
\beq
\label{SAbbratio}
    R^{\Upsilon(3S)-\Upsilon(1S)}|_{theory}&\simeq& \frac{.11 +.017}{2} 
\nonumber \\
          &\simeq& 0.06 \;,
\eeq  
in agreement with  the experimental ratio shown in Eq(\ref{3S1SPbpp}),
within experimental and theoretical errors.

 Similar results for $\Psi(2S)$ and $\Upsilon(3S)$ supression in p-Pb
collisions at an energy of 8 TeV can be found in Ref\cite{kdz16}

\section{Conclusions}

   The creation of a magnetic wall during the QCDPT, with B-B CMBR 
correlations is reviewed. This is a prediction for future cosmological
CMBR studies, and a test of the theory of the cosmological QCDPT. 

Then the use of QCD sum rules to show that the $\Psi(2S)$ and $\Upsilon(3S)$
are approximately 50\% normal $|q\bar{q}>$ states and 50\% hybrid  
$|q\bar{q}g>$ states was briefly reviewed.  
 
   The differential rapidity cross sections for
$J/\Psi, \Psi(2S)$ and $\Upsilon(nS)(n=1,2,3)$ production via p-p
collisions was reviewed. The enhancement of the production of $\Psi(2S)$ and 
$\Upsilon(3S)$ mixed heavy quark states was shown to be in agreement with
experiments. This is an important result for the test of the production
of QGP via RHIC.

  The differential rapidity cross sections for $\Psi$ and  $\Upsilon$
production via Cu-Cu and Au-Au collisions at RHIC (E=200 GeV) using 
$R^E_{AA}$, the product of the nuclear modification factor $R_{AA}$ and the 
dissociation factor $S_{\Phi}$, $N^{AA}_{bin}$ the binary collision number, and 
the gluon distribution function should give some guidance for future RHIC 
experiments. Then the possibly detection of the QGP via RHIC at the BNL
and LHC was reviewed.

  Finally, recent research on the suppression of heavy quark states, especially 
hybrid heavy quark states, produced via RHIC, as these mesons travel through 
nuclear matter was reviewed.
\newpage

{\bf Acknowledgements}

\normalsize
\vspace{5mm}

 The author acknowledges many helpful discussions with Dr. Debasish Das,
Saha Institute of Nuclear Physics, India, and support from the P25 group at 
Los Alamos National Laboratory.no

\end{document}